# Nanofluidic trapping and enhanced Raman detection of single biomolecules in plasmonic bowl-shaped nanopore

*Yingqi Zhao[1,2], Aliaksandr Hubarevich[1], Jian-An Huang[2]\*, and Francesco De Angelis[1]\**

[1]Istituto Italiano di Tecnologia, Via Morego 30, 16163, Genova, Italy. E-mail: francesco.deangelis@iit.it

[2]Faculty of Medicine, Faculty of Biochemistry and Molecular Medicine, University of Oulu, Aapistie 5 A, 90220 Oulu, Finland. E-mail: jianan.huang@oulu.fi

**ABSTRACT:** Solid-state nanopores are emerging platforms for single-molecule protein sequencing due to their tolerance to hash physiology environment and compatibility with different electrical and optical detection methods. However, they suffer from poor molecular manipulations that were twisted with and thus limited by the detection methods. Here, we report a bowl-shaped plasmonic gold nanopore on silicon nitride with hydrogel to demonstrate near-field nanofluidic manipulation of DNA translocation for plasmon-enhanced Raman spectroscopic detection. The hydrogel linearized the DNA, and the linear DNA was trapped in the nanopore for tens of seconds due assumably to bipolar effect of the nanopore that generate electroosmotic sheath flow and bipolar surface charge distribution. Their combination led to a near-field confinement of the DNA in the nanopore hot spot to allow stable Raman detection. We envision that a combination of Raman spectroscopy with the bowl-shaped nanopores can succeed in single-molecule protein sequencing in a label-free way.

## 1. Introduction

Nanopore sequencing capable of resistive pulse long-read of single DNA molecules in a portable device is playing an important role in personalized medicine and digital health.[1] To extend it to single-molecule protein sequencing, various types of solid-state nanopores are being developed to include optical and electrical detection methods as well as stability in different physiological environment.[2] In particular, plasmonic nanopores emerges because they not only achieved single-molecule detection of the 20 proteinogenic amino acids by surface-enhanced Raman spectroscopy (SERS) but also discriminated single residues in single peptides.[3]

As another pillar of single-molecule sequencing, molecular manipulation compatible to the solid-state nanopores that mimics function of nuclear acid movement by enzyme on biological nanopores are still challenging.[4] Electric bias in solid-state nanopores usually generates focused electric field in the nanopore to thread the DNA molecules through the nanopore and stretch them to ensure their single-file translocation. The latter requires so strong electric field in the solid-state nanopore that it drove the molecule through the nanopore too fast to be detected. For example, the translocation speed of a double-strand DNA molecule through a nanopore is around 30 basepairs per microsecond under a bias of a few hundreds of millivolts.[5] The case is even worse in sequencing single protein molecules due to uneven charge of amino acid residues.[6]

An enhanced electromagnetic field, termed hot spot, at a plasmonic gold bowtie nanopore was simulated to detect molecule sequence by SERS signals and generate strong optical forces able to trap single DNA molecules.[7] However, it was not realized experimentally probably due to the fast and complicated movement of the molecules inside the nanopore under electric bias. Despite strong optical force, the two plasmonic hot spots in the bowtie nanopore were so localized with small volumes as those of DNA bases that single molecules under Brownian movement had low probability to pass them through.

Methods to slow down the molecule translocation in plasmonic nanopores were developed including hydrogel nanopores, electro-osmosis flow control, and thermophresis.[8] For example, we demonstrated that agarose hydrogel could slowdown, unfold and stretch lambda DNA molecules in a porous gold nanopores such that SERS signals of their segments can be observed.[9] However, a near-field manipulation is still missing that traps the molecule into the small hot spots in the nanopore for a certain period for SERS detection. It could be the last piece of the puzzle for SERS sequencing of single protein molecules in plasmonic nanopore that has achieved identification of the 20 amino acids.

In this paper, we developed a bowl-shaped plasmonic nanopore on silicon nitride to combine the hydrogel slowdown of molecule translocation for SERS detection of lambda DNA, as shown in Figure 1A. Agarose hydrogel was placed below the nanopore to stretch the DNA molecules into linear chain before they were driven to the nanopore under electric bias. Upon laser excitation, the bowl shape structure of the nanopore (Figure 1B, C) focused optical energy on the nanopore edge to generate strong hot spots for SERS detection of the DNA molecule. The tapering metal thickness resulted in the narrow enhancement field at the nanopore edge therefore improve the special resolution of detection. Notably, long-time trappings of the DNA in the nanopore up to tens of seconds were observed repeatedly under bias of 30 mV. We assumed a bipolar electrode effect[10,11] on the nanopore that led to the trapping. When the molecule entered the nanopore, the nanopore under bias became a bipolar electrode with positive charge that attracted the negative-charged DNA molecules on the nanopore edge.[12] Our work suggests that a combination of the slowdown methods and near-field trapping techniques is needed for effective single-molecule sequencing in the plasmonic nanopores.

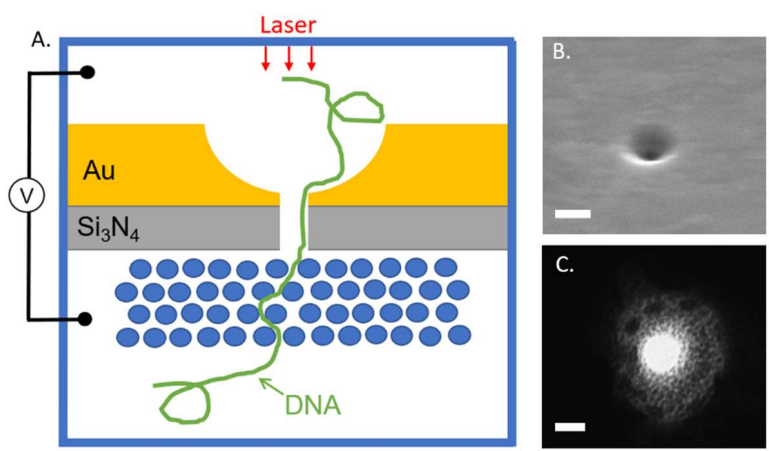

*Figure 1. (A)Schematic of the bowl-shaped plasmonic nanopore. (B) SEM image and (C) TEM image of the bowl-shaped plasmonic nanopore. The scale bar is 150nm and 20nm respectively.*

## 2. Results and discussion

The bowl-shaped nanopore was fabricated by Focused Ion Beam milling on the silicon nitride and sputtering deposition of gold film. The bowl shape was achieved by sculpturing the gold film with a group of concentric ring patterns. The sputtering depth decreased from the pattern center to the edge. The sputtering depths of each ring were tuned to form a bowl-shaped profile. After the bowl was sculptured, one ion beam shot was given to the bowl center point to penetrate the nitride and generated a nanopore. Then gold or gold-silver alloy film was sputtered on the front side of the nitride to further define the nanopore diameter. The nanopore was then encapsulated in a microfluidic chamber made from polydimethylsiloxane (PDMS). Agarose hydrogel solution was dropped into the low chamber by a pipette and then cool down according to our previous protoco[19]. Lambda DNAs in LiCl electrolyte were dropped into the lower chamber and translocated through the nanopore under electric bias across the nanopore and 785 nm laser illumination.

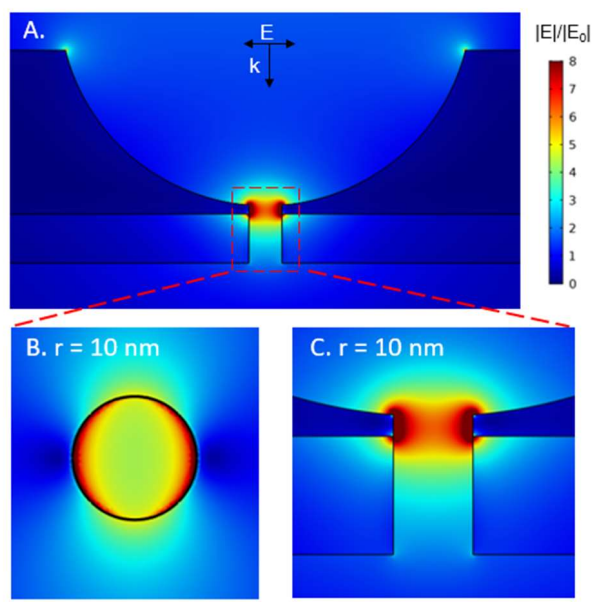

*Figure 2. The simulated electric field intensity distribution of the nanopore at wavelength 785 nm. (A) sideview, (B)magnified top view and (C) magnified sideview. The color bar indicates the electric field*

The bowl shape was aimed at confining and focusing illuminating laser energy at the nanopore edge for SERS measurements. Numerical simulation shows that optimal plasmonic resonance of the nanopores near the laser wavelength of 785 nm requires a nanopore diameter of around 20 nm. (Figure 2) The bowl shape here works to squeeze the 785 nm light into the small nanopore to generate strong electromagnetic field for SERS detection. Simulated field intensity distribution shows that incident light energy at wavelength 785 nm is confined and enhanced at the nanopore edge (Figure 2A-C). At the edge of tapering metal film, the narrow enhancement field enables detection of molecules with improved special resolution.

Monolayer of 4-ABT molecules were attached on the nanopore surface through thiol bonds for SERS performance validation. The nanopores with both 20 nm and 45 nm in diameter show distinct and stable time-lapsed SERS signals of 4-ABT (Figure 3A,B). Due to the narrow Raman peaks, the background SERS signals of 4-ABT were not overlap with those of DNA at range of 500 – 1000 cm$^{-1}$. The robust adsorbed 4-ABT monolayers on the nanopore surface also worked to prevent DNA from attaching on the nanopore surface.

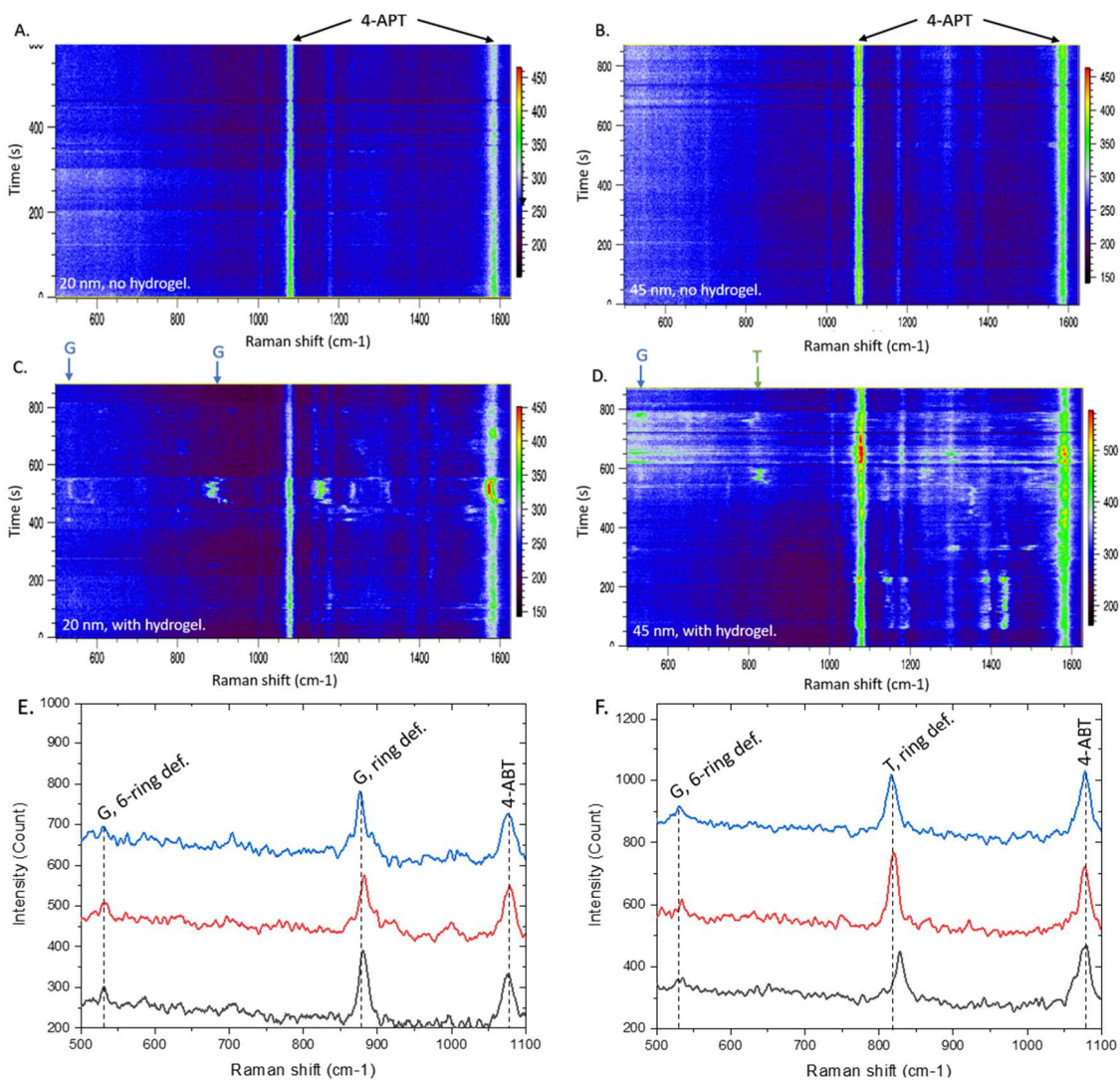

*Figure 3. Time trace of Raman spectra of lambda-DNA that pass through the bowl-shaped nanopores under electric bias of 30 mV in 1 M LiCl: (A)20 nm diameter nanopore without hydrogel. (B) 45 nm diameter nanopore without hydrogel. (C) 20 nm diameter nanopore with hydrogel. (D) 45 nm diameter nanopore with hydrogel. (E) Typical Rama spectra extracted from (C) during 500 – 550 s. (F) Typical Rama spectra extracted from (D) during 550 – 600 s; def. means deformation.*[13]

We observed molecular trapping effect when electric bias of 30 mV was used to drive the lambda-DNA through the 4-ABT-coverred nanopores. In case of no hydrogel slowdown, no SERS signals of DNA was observed, as shown in Figure 2A,B. When hydrogel was used, stable and continuous SERS signals of Guanine (G) were observed in the 20 nm nanopore for around 75 s from 485 to 550 s (Figure 2C).[13] We observed similar phenomena in the 45 nm nanopore with hydrogel slowdown. Stable Guanine and Thymine (T) signals appeared continuously for around 70 s, followed by signals fluctuation with time (Figure 2D). The latter implied conformation changes of the DNA segment from the linear chain.[3] Similar stable and continuous SERS spectra of the DNA segments were found repeatably in our measurements by different bias voltages.

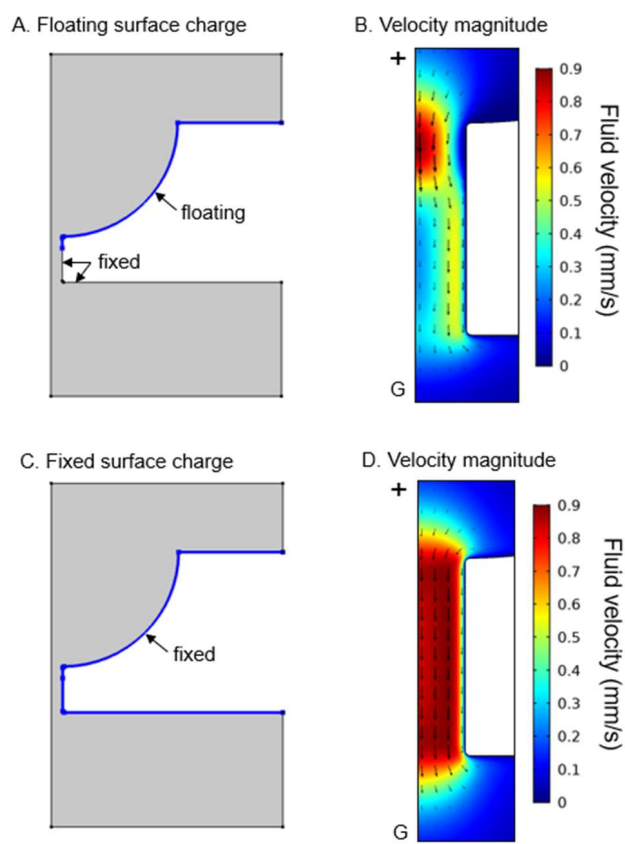

*Figure 4. Multiphysics simulation of the biopolar trapping effect and electroosmic flow in the bow-shaped nanopore. (A) Model of a gold bowl-shaped nanopore with fixed surface potential and its (B) simulated electroosmic sheath flow in the nanopore. (C) Model of a gold bowl-shaped nanopore with floating surface potential and its (D) simulated uniform electroosmic flow in the nanopore. The colour bar indicates the fluid velocity.*

The continuous peaks suggest that a linear segment of the lambda DNA stayed in the hot spot for the periods. The linear confirmation of the lambda DNA chain in the hot spot have been reported in previous literatures and our previous work that used the hydrogel to stretch the DNA.[9,14] Otherwise, peak change should have been observed.[3] Furthermore, while the 4-ABT peak positions remain stable in both Figure 3E and F, the positions of the Guanine ring deformation peak (Figure 3E) and the Thymine ring deformation peak (Figure 3F) fluctuate. Such fluctuations were likely assigned to diffusion of the segment of single DNA chain.[3,15] In particular, no adenine peaks were observed, which ruled out the possibility that the DNA segment attached on the nanopore surface by the adenine interaction with gold.[16] All these evidence suggest a dynamic near-field trapping of the linear DNA segment in the hot spot.

As DNA molecule should pass through nanopore rapidly, our finding shown that the molecules were trapping in the hot spots for ten of seconds. We assume the bipolar electrode effect of the nanopore on silicon nitride as the trapping mechanism that generate floating positive surface potential under the electric bias.[11] The nanopore had floating potential while the nitride part had fixed potential (Figure 4A). As a result, electroosmotic sheath flow in the downward direction was generated in the nanopore where fast flow was in the nanopore center and slow flow occurred near the nanopore sidewall (Figure 4B). In contrast, a bowl-shaped nanopore fully wrapped with gold layer (Figure 4C) with a fixed potential throughout the whole nanopore area would not have the bipolar effect. The corresponding electroosmotic flow was uniform in the nanopore (Figure 4D).

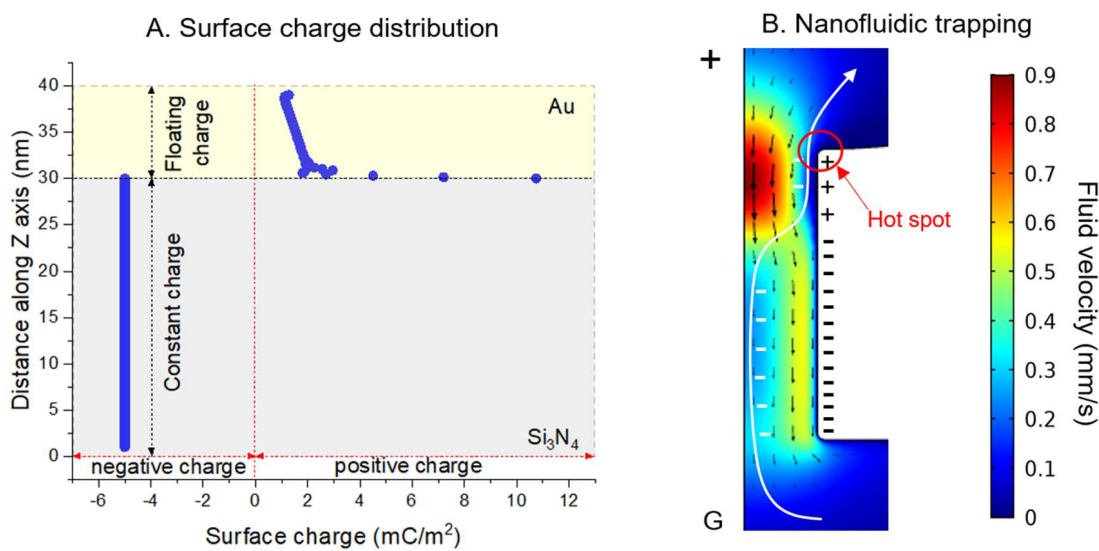

*Figure 5. (A) Multiphysics simulation of the surface charge in the bowl-shaped nanopore on silicon nitride. (B) Nanofluidic trapping mechanism. The color bar indicates the fluid velocity.*

In addition, the bipolar effect led to positive surface charge of the gold sidewall but negative surface charge on the nitride side wall in the nanopore (Figure 5A). The bipolar charge distribution would drive the negatively-charged DNA chain to the slow downward flow regions that were actually favorable for the upwards moving DNA (Figure 5B). A combination of these two effect actually prevented the linear DNA from Brownian movement in the nanopore and attracted the DNA to the hot spot of the nanopore edge.[12] Due to the surface protection by the 4-ABT monolayer in the hot spot, such near-field nanofluidic trapping allowed the DNA stayed in the hot spot for tens of seconds and emit stable SERS spectra.

## 3. Conclusion

To summarize, we integrated a hydrogel slowdown in a bowl-shaped nanopore for SERS detection of lambda-DNA molecules in LiCl electrolyte under electric bias. The stable and continuous SERS spectra demonstrated a nanofluidic trapping of the linear DNA in the nanopore for tens of second. The trapping mechanism was investigated by multiphysics simulation of the bipolar effect of the bowl-shaped nanopore that generate electroosmotic sheath flow as well as bipolar surface charge distribution. Their combination led to a confinement of the DNA on the nanopore hot spot that generate the stable SERS signals. Our finding demonstrated an effective near-field molecular trapping method that, combined with the hydrogel slowdown method, has high potentials for plasmonic nanopores in single-molecule analysis and sequencing.

## 4. Author contributions

Yingqi Zhao fabricated the nanopore samples and did the Raman measurements. Aliaksandr Hubarevich did the Multiphysics simulation and investigated the trapping mechanism. Jian-An Huang helped with Raman measurements, drafted the manuscript, investigated the trapping mechanism and supervised the work. Francesco De Angelis designed the research, investigated the trapping mechanism and supervised the work.

## 5. Acknowledgement

This work has received funding from the European Union's Horizon 2020 research and innovation programme through the project PROID under grant agreement No. 964363. Jian-An Huang acknowledges the DigiHealth-project, a strategic profiling project at the University of Oulu that is supported by the Academy of Finland (project number 326291) and the University of Oulu.